\documentclass[english,letterpaper]{revtex4}
\usepackage[T1]{fontenc}
\usepackage[latin1]{inputenc}
\usepackage{float}
\usepackage{graphicx}
\usepackage{amssymb}

\makeatletter

\providecommand{\tabularnewline}{\\}

\usepackage{float}

\makeatletter

\usepackage{float}

\makeatletter

\usepackage{float}

\makeatletter

\usepackage{float}

\makeatletter

\usepackage{geometry}

\makeatletter

\usepackage{dcolumn}

\usepackage{bm}

\makeatother

\makeatother

\makeatother

\makeatother

\makeatother

\usepackage{babel}
\makeatother
\begin{document}

\title{Relativistic multi-reference Fock-space coupled-cluster calculation
of the forbidden $6s^{2}\,^{1}S_{0}\longrightarrow6s5d\,^{3}D_{1}$
magnetic-dipole transition in ytterbium}

\author{Chiranjib Sur}

\affiliation{Department of Astronomy, The Ohio State University Columbus, Ohio,
43210, USA}

\email{csur@astronomy.ohio-state.edu}

\homepage{http://www.astronomy.ohio-state.edu/~csur}

\author{Rajat K Chaudhuri}

\affiliation{Indian Institute of Astrophysics,Bangalore 560034, India}

\date{Phys. Rev. A, \textbf{76}, 012509 (2007)}

\begin{abstract}
We report the forbidden $6s^{2}\,^{1}S_{0}\longrightarrow6s5d\,^{3}D_{1}$
magnetic-dipole transition amplitude computed using multi-reference
Fock-space coupled-cluster theory. Our computed transition matrix
element ($1.34\times10^{-4}\mu_{B}$) is in excellent agreement with
the experimental value ($1.33\times10^{-4}$ $\mu_{B}$). This value
in combination with other known quantities will be helpful to determine
the parity non-conserving amplitude for the $6s^{2}\,^{1}S_{0}\longrightarrow6s5d\,^{3}D_{1}$
transition in atomic Yb. To our knowledge our calculation is the most
accurate to date and can be very important in the search of physics
beyond the standard model. We further report the $6s6p\,^{3}P_{0}\longrightarrow6s6p\,^{1}P_{1}$
and $6s5d\,^{3}D_{1}\longrightarrow6s6p\,^{3}P_{0}$ transition matrix
elements which are also in good agreement with the earlier theoretical
estimates.

~

\textbf{PACS number(s)} : 31.15.Ar, 31.15.Dv, 31.25.-v, 32.70.Cs, 
\end{abstract}
\maketitle

\section{\label{intro}Introduction}

The highly forbidden $6s^{2}\,^{1}S_{0}\longrightarrow6s5d\,\,^{3}D_{1}$
magnetic-dipole ($M1$) transition amplitude in ytterbium (Yb), a
key quantity for evaluating the feasibility of parity non-conservation
(PNC), has recently been measured by Stalnaker \emph{et al.} \cite{Budker}
using Stark-interference experiment. The electric-dipole ($E1$) matrix
element for $6s^{2}\,^{1}S_{0}\longrightarrow6s5d\,\,^{3}D_{1}$ transition
in Yb is forbidden because of its $s-d$ nature. The forbidden $M1$
transition amplitude mentioned above is therefore the key quantity
to explore the feasibility of the PNC study for this transition in
Yb. Accurate determination of the $M1$ transition amplitude, which
is strongly suppressed in nature in the absence of external fields,
can be used together with the large PNC- and moderately large Stark-
induced $E1$ amplitudes to understand PNC studies in neutral Yb.
Strong configuration mixing and spin-orbit interaction in both the
upper and the lower states give rise to a non-zero $6s^{2}\,^{1}S_{0}\longrightarrow6s5d\,\,^{3}D_{1}$
transition amplitude \cite{Budker,Hw}. Surprisingly, despite its
tremendous importance in PNC experiments, only a rough theoretical
estimate ($|A(M1)|\leqslant10^{-4}\mu_{B}$) is available in the literature
for this transition. PNC in atoms arises from the neutral weak interactions
and are considerably enhanced in heavy atoms. Combining the high precision
experiments and theoretical calculations of PNC observables, it is
possible to extract the nuclear weak charge \cite{PNC-review}. Any
discrepancy of its value with the one obtained from the standard model
(SM) of particle physics could possibly reveal the existence of new
physics beyond the SM.

\begin{figure}
\begin{centering}
\includegraphics[scale=0.7]{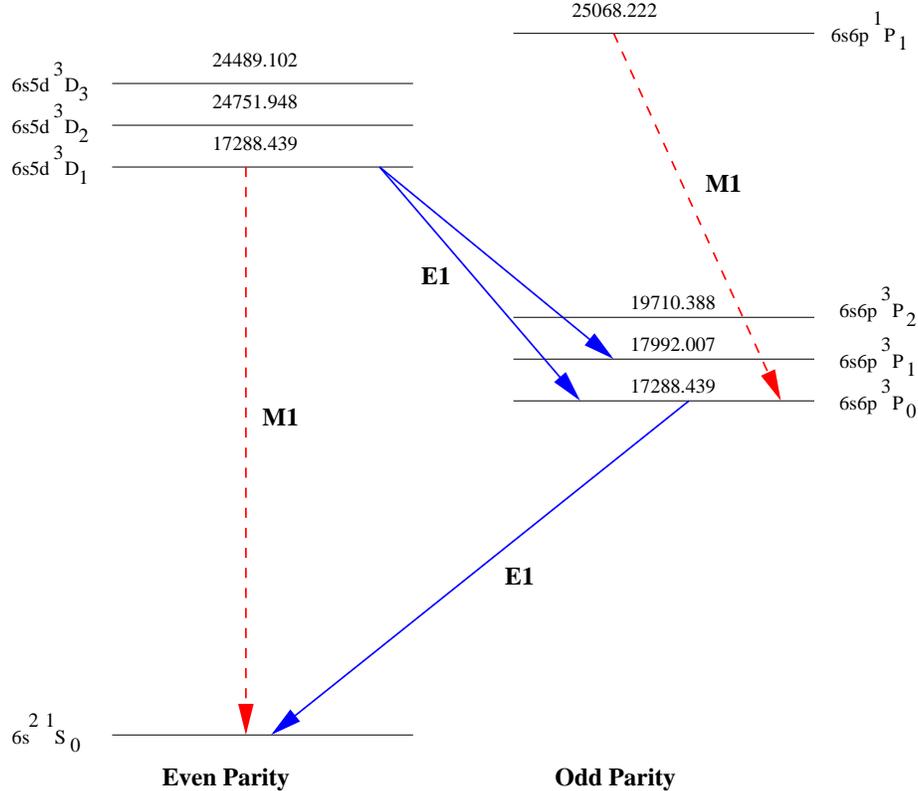} 
\par\end{centering}

\caption{\label{Yb-levels}Energy levels of the ground and low lying excited
states of Yb. The energies (in $\mathrm{cm}^{-1}$) are given with
respect to the ground state and are obtained from the NIST database
\cite{NIST}. Electric dipole (allowed) and magnetic dipole (forbidden)
transitions are represented by `blue' (solid) and `red' (dashed) lines
respectively. This diagram helps us to understand the requirement
of a multi-reference theory to describe the atomic states of Yb.}
\end{figure}

The ground and excited states of closed shell ground state systems
like Yb are, in general, multi-configurational in nature, and hence,
an accurate description of these states requires a balanced treatment
of non-dynamical or configuration mixing and dynamical electron correlation
effects (this will be more clear by studying the energy levels in
figure \ref{Yb-levels}). It is, therefore, imperative that these
systems must be treated with methods which are combinations of the
configuration interaction (CI) and many-body perturbation theory (MBPT),
such as multi-reference many-body perturbation theories (MR-MBPT)
\cite{Hv,MRMP,PT2,Hvr,Nakano,Dzuba}, multi-reference Fock-space coupled-cluster
(MR-FSCC) theories and/or it variants \cite{mrcc,Lindgren,Mukherjee,Kaldor,Pie}
etc. The state-of-the-art MR-FSCC is an all-order approach and is
capable of providing reliable estimates of predicted quantities. In
this paper, we employ the MR-FSCC method to compute the magnetic-dipole
transition amplitude for $6s^{2}\,^{1}S_{0}\longrightarrow6s5d\,\,^{3}D_{1}$
transition in Yb using four-component relativistic spinors. The resulting
value of this magnetic-dipole transition matrix element in atomic
Yb is $1.34\times10^{-4}\mu_{B}$, which differs by less than one
percent from the experimental value. In addition, we have also calculated
the $6s6p\,^{3}P_{0}\longrightarrow6s6p\,^{1}P_{1}$ $M1$ transition
transition amplitude in Yb which plays crucial role in the measurement
of the PNC induced electric-dipole amplitude \cite{Kimball}. This
is the first time any variant of coupled-cluster theory has been applied
to determine the $M1$ transition amplitude of Yb. A precise determination
of this quantity ensures not only the power of the theory but also
for the experimental uncertainties. To our knowledge no such theoretical
results are available for magnetic dipole transitions in Yb.

The structure of this paper is the following : section \ref{intro}
describes the physical relevance of the problem. Section \ref{OSCC}
provides a brief outline of the multi-reference Fock-space CC (MR-FSCC)
theory for two-electron attachment processes that is used to compute
the M1 transition elements between the ground $^{1}S_{0}$ and excited
$^{3}D_{1}$ state. Section \ref{res} contains the results of our
calculation with an in-depth discussion. Finally in section \ref{conclusion}
we conclude and highlight the findings of our paper.

\section{\label{OSCC}Fock-space multi-reference coupled-cluster (mr-fscc)
theory for two-electron attachment processes}

\begin{figure}[H]
\begin{centering}
\includegraphics[scale=0.5]{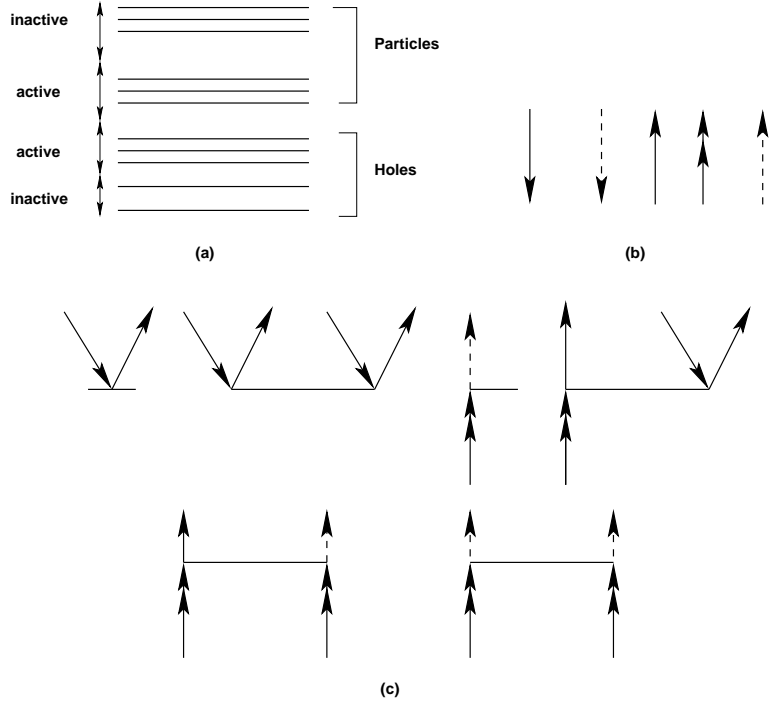} 
\par\end{centering}

\caption{\label{fig-One}(a) Schematic depiction of the classification of
particle and hole orbitals into active and inactive categories. (b)
Diagrammatic representation of hole ($\downarrow$), particles ($\uparrow$),
active particles (double up arrow), inactive holes, and particles
(dashed down/up arrow). (c) Diagrammatic representation of $S^{(0,0)}$
$(T)$, $S^{(0,1)}$, and $S^{(0,2)}$ cluster operators.}
\end{figure}

In MR-FSCC method \cite{Lindgren,Haque,Mukherjee,LindMukh,Sinha,Pal,Kaldor},
the self-consistent field (SCF) solution of the Hartree-Fock (Dirac-
Fock in relativistic regime) for the $N$-electron closed shell ground
state $\Phi_{\mathrm{HF/DF}}$ is chosen as the vacuum (for labeling
purpose only) to define holes and particles with respect to $\Phi_{\mathrm{HF/DF}}$.
The multi-reference aspect is then introduced by subdividing the hole
and particle orbitals into active and inactive categories, where different
occupations of the active orbitals will define a multi-reference \emph{model}
space for our problem. We call a model space to be \emph{complete}
if it has all possible electron occupancies in the active orbitals,
otherwise incomplete. The classification of orbitals into active and
inactive groups is, \emph{in principle}, arbitrary and is at our disposal.
However, for the sake of computational convenience, we treat only
a few hole and particle orbitals as active, namely those are close
to the Fermi level. The classification of orbitals is depicted schematically
in Fig.\ref{fig-One}(a). Diagrammatically, active holes and particles
are depicted as solid lines with double arrows and the corresponding
inactive lines are designated by dotted lines with single arrow. The
orbitals which can be \emph{both} active and inactive are designated
by solid lines with single arrow (see Fig.\ref{fig-One}(b)).

We designate by $\Psi_{i}^{0(k,l)}$ a model space of $k$-hole and
$l$-particle determinants, where in the present instance ($\mathrm{Yb}^{+2}+2e\longrightarrow\mathrm{Yb}$),
$k=0$ and $l$ ranges from 0 to 2. Generally, any second quantized
operator has $k$-hole and $l$-particle annihilation operators for
the active holes and particles. For convenience, we indicate the {}``hole-particle
valence rank'' of an operator by a superscript ($k,l$) on the operator.
Thus, according to our notation, an operator $A^{(k,l)}$ will have
exactly $k$-hole and $l$-particle annihilation operators.

We now describe the type of ansatz used to derive the MR-FSCC equations
for direct energy difference calculations in two-electron attachment
processes. The Hartree-Fock/Dirac-Fock function $\Phi_{\mathrm{HF/DF}}$
is denoted by $\Psi^{(0,0)}$ and the inactive hole and particle orbitals
(defined with respect to $\Phi_{\mathrm{HF/DF}}$) are labeled by
the indices $a,b,c,\cdots$ and $p,q,r,\cdots$, respectively. The
corresponding active holes and particles are labeled by the indices
$\alpha,\beta,\gamma\cdots$ and $u,v,w\cdots$, respectively. Note
that there will be \emph{no active} holes (particles) for two electron
attachment (detachment) processes. The cluster operator correlating
the $N$-electron ground/reference state is denoted in our notation
by $S^{(0,0)}$ which can be split into various $n$-body components
depending upon the various hole-particle excitation ranks. The cluster
operator $S^{(0,0)}$ upto 2-body (first two diagrams of Fig.\ref{fig-One}(c))
can be written in second quantized notation as, \begin{equation}
S^{(0,0)}=S_{1}^{(0,0)}+S_{2}^{(0,0)}+\cdots=\sum_{p,a}\langle p|s_{1}^{(0,0)}|a\rangle{\{ a_{p}^{\dagger}a_{a}}\}+\frac{1}{4}\sum_{a,b,p,q}\langle pq|s_{2}^{(0,0)}|ab\rangle{\{ a_{p}^{\dagger}a_{q}^{\dagger}a_{b}a_{a}}\}+\cdots\label{eq1}\end{equation}
 where $a^{\dagger}$ ($a$) denotes creation (annihilation) operator
with respect to $\Phi_{\mathrm{HF/DF}}$ and ${\{\cdots}\}$ denotes
\emph{normal} ordering. It should be noted that $S^{(0,0)}$ cannot
destroy any holes or particles; acting of $\Phi_{\mathrm{HF/DF}}$,
it can only create them.

For $(N+1)$ electron states the model space consists of zero active
hole and one active particle ($k=0,l=1$) and hence according to our
notation the valence sector for $(N+1)$ electron states can be written
as (0,1) sector. We introduce an wave operator $\Omega$ which generates
all valid excitation from the model space function for $(N+1)$ electron
states. The wave operator $\Omega$ for the (0,1) valence problem
is given by \begin{equation}
\Omega={\{\exp(S^{(0,0)}+S^{(0,1)})}\}.\label{eq2}\end{equation}

In this case the additional cluster operator $S^{(0,1)}$ must be
able to destroy active particle present in the (0,1) valence space.
Like $S^{(0,0)}$, the cluster operator $S^{(0,1)}$ can also be split
into various $n$-body components depending upon hole-particle excitation
ranks. The one- and two-body $S^{(0,1)}$ (3rd and 4th diagram of
Fig.\ref{fig-One}(c)) can be written in the second quantized notation
as \begin{equation}
S^{(0,1)}=S_{1}^{(0,1)}+S_{2}^{(0,1)}+\cdots=\sum_{p\ne u}\langle p|s_{1}^{(0,1)}|u\rangle{\{ a_{p}^{\dagger}a_{u}}\}+\frac{1}{2}\sum_{p,q,a}\langle pq|s_{2}^{(0,1)}|ua\rangle{\{ a_{p}^{\dagger}a_{q}^{\dagger}a_{b}a_{u}}\}+\cdots\label{eq3}\end{equation}
 where $u$ denotes the active particle which is destroyed.

Similarly, for $(N+2)$ electron states (two-electron attachment processes)
the model space consists of zero active hole and two active particles
($k=0,l=2$) and the valence sector may be written as (0,2) sector.
In this case, the additional cluster operator must be able to destroy
two active particles and this may be designated by $S^{(0,2)}$. The
total wave operator $\Omega$ for the (0,2) problem is then given
by \begin{equation}
\Omega={\{\exp(S^{(0,0)}+S^{(0,1)}+S^{(0,2)})}\}.\label{eq4}\end{equation}
 A typical two-body $S_{2}^{(0,2)}$ operator (5th and 6th diagrams
of Fig.\ref{fig-One} (c)) may be written as \begin{equation}
S_{2}^{(0,2)}=\frac{1}{2}\sum_{p,q,u,v}\langle pq|s_{2}^{(0,2)}|uv\rangle{\{ a_{p}^{\dagger}a_{q}^{\dagger}a_{v}a_{u}}\},\label{eq5}\end{equation}
where $u$ and $v$ denote active particle which are destroyed. Note
that orbitals $p$ and $q$ both cannot be active at the same time.
We further emphasize that under two-body truncation scheme $S^{(0,2)}$=0,
\emph{if} all the particle orbitals are active.

In general, for a $(k,l)$ valence problem, the cluster operator must
be able to destroy any subset of $k$- active holes and $l$- active
particles. Hence, the wave operator $\Omega$ for $(k,l)$ valence
sector may be written as \begin{equation}
\Omega={\{\exp({\tilde{S}}^{(k,l)})}\},\label{eq6}\end{equation}
 where \begin{equation}
{\tilde{S}}^{(k,l)}={\sum_{m=0}^{k}}\hspace{0.1in}{\sum_{n=0}^{l}}S^{(m,n)}.\label{eq7}\end{equation}

To compute the ground to excited state transition energies and M1
transition element(s) of Yb, we begin with the Dirac-Coulomb Hamiltonian
($H$) for an $N$-electron atom which can be written as \begin{equation}
H=\sum_{i=1}^{N}\left[c\vec{\alpha_{i}}\cdot\vec{p}_{i}+\beta mc^{2}+V_{\mathrm{Nuc}}(r_{i})\right]+\sum_{i<j}^{N}\frac{e^{2}}{r_{ij}}\label{eq00}\end{equation}
 with all the standard notations often used. The normal ordered form
of the above Hamiltonian, relative to the mean field energy, is given
by \begin{equation}
\mathcal{H}=H-\langle\Phi|H|\Phi\rangle=H-E_{DF}=\sum_{ij}\langle i|\mathsf{\mathtt{f}}|j\rangle\left\{ a_{i}^{\dagger}a_{j}\right\} +\frac{1}{4}\sum_{i,j,k,l}\langle ij||kl\rangle\left\{ a_{i}^{\dagger}a_{j}^{\dagger}a_{l}a_{k}\right\} .\label{eq01}\end{equation}
 Here \begin{equation}
\langle ij||kl\rangle=\langle ij|\frac{1}{r_{12}}|kl\rangle-\langle ij|\frac{1}{r_{12}}|lk\rangle,\label{eq02}\end{equation}
 $E_{DF}$ is the Dirac-Fock energy and $\mathsf{\mathtt{f}}$ is
the one-electron Fock operator.

We define the exact wave function $\Psi_{i}^{(k,l)}$ for ($k,l$)
valence sector as \begin{equation}
\Psi_{i}^{(k,l)}=\Omega\Psi_{i}^{0(k,l)}\label{eq10}\end{equation}
 where \begin{equation}
\Psi_{i}^{0(k,l)}=\sum_{i}C_{i}^{(k,l)}\Phi_{i}^{(k,l)}.\label{eq11}\end{equation}
 The functions $\Phi_{i}^{(k,l)}$ in Eq.(\ref{eq11}) are the determinants
included in the model space $\Psi_{i}^{0(k,l)}$ and $C^{(k,l)}$
are the corresponding coefficients. Substituting the above form of
the wave-function (given in Eqs. (\ref{eq10}) and (\ref{eq11}))
in the Schr{ö}dinger equation for a manifold of states $H|\Psi_{i}^{(k,l)}\rangle=E_{i}|\Psi_{i}^{(k,l)}\rangle$,
we get \begin{equation}
H\Omega{\left(\sum_{i}C_{i}|\Phi_{i}^{(k,l)}\rangle\right)}=E_{i}\Omega{\left(\sum_{i}C_{i}|\Phi_{i}^{(k,l)}\rangle\right)},\label{eq9}\end{equation}
 where $E_{i}$ is the $i$-th state energy.

Following Lindgren \cite{Lindgren}, Mukherjee \cite{Mukherjee},
Lindgren and Mukherjee \cite{LindMukh}, Sinha \emph{et al.} \cite{Sinha}
and Pal \emph{et al.} \cite{Pal}, the Fock-space Bloch equation for
the MR-FSCC may be written as \begin{equation}
H\Omega P^{(k,l)}=P^{(k,l)}H_{\mathrm{eff}}^{(k,l)}\Omega P^{(k,l)}\hspace{0.2in}\forall(k,l),\label{eq12}\end{equation}
 where \begin{equation}
H_{\mathrm{eff}}^{(k,l)}=P^{(k,l)}\Omega^{-1}H\Omega P^{(k,l)}\label{eq13}\end{equation}
 and $P^{(k,l)}$ is the model space projection operator for the ($k,l$)
valence sector (defined by $\sum_{i}C_{i}^{(k,l)}\Phi_{i}^{(k,l)}$).
For complete model space, the model space projector $P^{(k,l)}$ satisfies
the \emph{intermediate} normalization condition \begin{equation}
P^{(k,l)}\Omega P^{(k,l)}=P^{(k,l)}.\label{eq14}\end{equation}

\begin{figure}[H]
\begin{centering}
\includegraphics[scale=0.5]{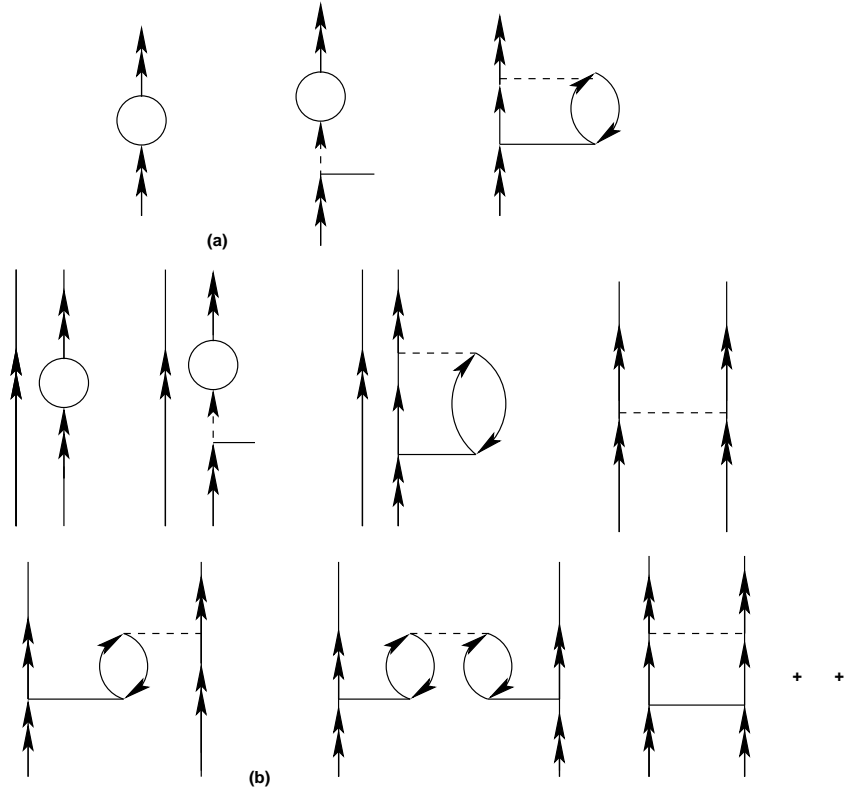} 
\par\end{centering}

\caption{\label{fig-TWO}Diagrammatic representation of $H_{\mathrm{eff}}^{(0,1)}$(figure
a) and $H_{\mathrm{eff}}^{(0,2)}$(figure b). The one- and two-body
dressed operators of $\widetilde{H}$ are represented by circle and
dashed lines, respectively. Exchange diagrams are not shown here for
convenience.}
\end{figure}

At this juncture, we single out the cluster amplitudes $S^{(0,0)}$
and call them $T$. The rest of the cluster amplitudes will henceforth
be called $S$ and are shown in Fig. \ref{fig-One}. The normal ordered
definition of $\Omega$ enables us to rewrite Eq.(\ref{eq7}) as

\begin{equation}
\Omega=\exp(T)\{\exp(S)\}=\exp(T)\Omega_{v}\label{eq15}\end{equation}
 where $\Omega_{v}$ represents the wave-operator for the valence
sector.

To formulate the theory for direct energy differences, we pre-multiply
Eq.(\ref{eq12}) by $\exp(-T)$ and get \begin{equation}
\overline{H}\Omega_{v}P^{(k,l)}=\Omega_{v}P^{(k,l)}H_{\mathrm{eff}}^{(k,l)}P^{(k,l)}\ ,\,\,\,\,\,\forall(k,l)\ne(0,0)\label{eq16}\end{equation}
 where $\overline{H}=\exp(-T){H}\exp(T)$. Since $\overline{H}$ can
be partitioned into a connected operator $\widetilde{H}$ and $E_{\mathrm{ref/gr}}$
($N$-electron closed-shell reference or ground state energy), we
likewise define $\widetilde{H}_{\mathrm{eff}}$ as \@.\begin{equation}
\widetilde{H}_{\mathrm{eff}}^{(k,l)}=H_{\mathrm{eff}}^{(k,l)}-E_{\mathrm{gr}},\,\,\,\,\,\forall(k,l)\ne(0,0).\label{eq17}\end{equation}

\begin{figure}[H]
\begin{centering}
\includegraphics[scale=0.6]{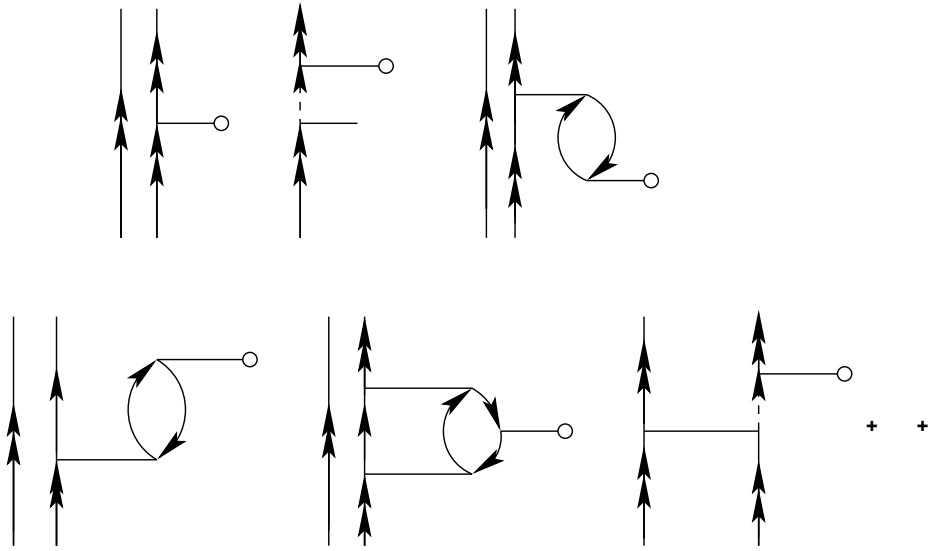} 
\par\end{centering}

\caption{\label{fig-Three}Diagrammatic representation of $\langle\Psi_{f}^{(0,2)}|M1|\Psi_{i}^{(0,2)}\rangle$.
The one-body dressed operator $\overline{M1}=\exp(T^{\dagger})M1\exp(T)$
is represented by line with circle. Exchange diagrams are not shown
here for convenience.}
\end{figure}

Substituting Eq.(\ref{eq17}) in Eq.(\ref{eq16}) we obtain the Fock-space
Bloch equation for energy differences: \begin{equation}
\widetilde{H}\Omega_{v}P^{(k,l)}=\Omega_{v}P^{(k,l)}\widetilde{H}_{\mathrm{eff}}^{(k,l)}P^{(k,l)},\,\,\,\,\,\forall(k,l)\ne(0,0).\label{eq18}\end{equation}
Eqs. (\ref{eq12}) and (\ref{eq18}) are solved by the Bloch projection
method for $k=l=0$ and $k=0,l\ne0$, respectively, involving the
left projection of the equations with $P^{(k,l)}$ and its orthogonal
complement $Q^{(k,l)}$ ($P^{(k,l)}+Q^{(k,l)}$=1) to obtain the effective
Hamiltonian and the cluster amplitudes, respectively. At this point,
we recall that the cluster amplitudes in MR-FSCC are solved hierarchically
through the \emph{subsystem embedding condition} (SEC) \cite{SEC,Haque}
which is equivalent to the \emph{valence universality} condition used
by Lindgren \cite{Lindgren} in his formulation. For example, in the
present application, we first solve the MR-FSCC for $k=l=0$ to obtain
the cluster amplitudes $T$. The operator $\widetilde{H}$ and $\widetilde{H}_{\mathrm{eff}}^{(0,1)}$
are then constructed from this cluster amplitudes $T$ to solve Eq.
(\ref{eq18}) for $k=0$, $l=1$ to determine $S^{(0,1)}$ amplitudes.
The effective Hamiltonian for $(0,1)$ Fock space (represented diagrammatically
in Fig.\ref{fig-TWO}), constructed from ${H}$, $T$ and $S^{(0,1)}$
is then diagonalized within the model space to obtained the desired
eigenvalues and eigenvectors. The diagonalization is followed from
the eigenvalue equation

\begin{equation}
\widetilde{H}_{\mathrm{eff}}^{(0,1)}C^{(0,1)}=C^{(0,1)}E,\label{eq19}\end{equation}
 where \begin{equation}
\widetilde{H}_{\mathrm{eff}}^{(0,1)}=P^{(0,1)}[\widetilde{H}+\overbrace{\widetilde{H}S^{(0,1)}}]P^{(0,1)}.\label{eq20}\end{equation}
 The expression $\overbrace{\widetilde{H}S^{(0,1)}}$ in Eq.(\ref{eq20})
indicates that operators $\widetilde{H}$ and $S^{(0,1)}$) are connected
by common orbital(s).

The MR-FSCC equations for $(0,2)$ sector are then solved to determine
$S^{(0,2)}$ where the cluster amplitudes from the lower valence sectors
behave as {}``known quantities''. The effective Hamiltonian for
the $(0,2)$ Fock space constructed from $H$, $T$, $S^{(0,1)}$
and $S^{(0,2)}$ is then diagonalized to get the desired roots by
using the equation \begin{equation}
\widetilde{H}_{\mathrm{eff}}^{(0,2)}C^{(0,2)}=C^{(0,2)}E\,.\label{eq21}\end{equation}
 where \begin{equation}
\widetilde{H}_{\mathrm{eff}}^{(0,2)}=P^{(0,2)}[\widetilde{H}+\overbrace{\widetilde{H}S^{(0,1)}}+\frac{1}{2}\overbrace{\widetilde{H}S^{(0,1)}S^{(0,1)}}+\overbrace{\widetilde{H}S^{(0,2)}}]P^{(0,2)}.\label{eq22}\end{equation}
 It is worth noting that the eigenvalue and eigenfunctions for the
$(0,1)$ valence sector are by-products of MR-FSCC for the $(0,2)$
valence sector with no additional computation. Once the cluster amplitudes
are known, the magnetic-dipole matrix element between the two states
can be computed using the following expression \begin{eqnarray}
\langle\mathrm{Final\, state}(f)|M1|\mathrm{Initial\, state}(i)\rangle & = & \frac{\langle\Psi_{f}^{(0,2)}|M1|\Psi_{i}^{(0,2)}\rangle}{\sqrt{\langle\Psi_{f}^{(0,2)}|\Psi_{f}^{(0,2)}\rangle\langle\Psi_{i}^{(0,2)}|\Psi_{i}^{(0,2)}\rangle}}\label{eq21}\end{eqnarray}
 where $|\Psi_{i}^{(0,2)}\rangle$ and $|\Psi_{f}^{(0,2)}\rangle$
are the exact initial and final states, respectively. With aid of
$\Omega$, the valence universal wave operator, Eq.(\ref{eq16}) can
be further simplified to \begin{eqnarray}
\langle M1\rangle_{fi} & = & \frac{\langle\Phi_{f}^{(0,2)}|(1+S^{\dagger}){\overline{M1}}(1+S)|\Phi_{i}^{(0,2)}\rangle}{\sqrt{\langle\Phi_{f}^{(0,2)}|(1+S^{\dagger})e^{T^{\dagger}}e^{T}(1+S)|\Phi_{f}^{(0,2)}\rangle\langle\Phi_{i}^{(0,2)}|(1+S^{\dagger})e^{T^{\dagger}}e^{T}(1+S)|\Phi_{i}^{(0,2)}\rangle}}\,,\label{M1-me}\end{eqnarray}
 where $\overline{M1}=\exp(T^{\dagger})M1\exp(T)$ and $S=S^{(0,1)}+S^{(0,2)}$.

The single particle reduced matrix elements for the $M1$ transition
is given by,

\begin{equation}
\left\langle \kappa_{f}\right\Vert m1\left\Vert \kappa_{i}\right\rangle =\frac{6}{\alpha k}\left\langle j_{f}\right\Vert \mathcal{C}_{q}^{(1)}\left\Vert j_{i}\right\rangle \times\left(\frac{\kappa_{f}+\kappa_{i}}{2}\right)\int\mathtt{j}_{1}(kr)\left(\mathcal{P}_{f}\mathcal{Q}_{i}+\mathcal{Q}_{f}\mathcal{P}_{i}\right)dr.\label{m1-mat}\end{equation}
 Here $j$'s and $\kappa$'s are the total orbital angular momentum
and the relativistic angular momentum quantum numbers respectively;
$k$ is defined as $\omega\alpha$ where $\omega$ is the single particle
difference energy and $\alpha$ is the fine structure constant. The
single particle orbitals are expressed in terms of the Dirac spinors
with $\mathcal{P}_{i}$ and $\mathcal{Q}_{i}$ as the large and small
components for the $i$th spinor, respectively. The angular coefficients
are the reduced matrix elements of the spherical tensor of rank $m$
and are expressed as

\begin{equation}
\left\langle \kappa_{f}\right\Vert \mathcal{C}_{q}^{(m)}\left\Vert \kappa_{i}\right\rangle =(-1)^{j_{f}+1/2}\sqrt{(2j_{f}+1)(2j_{i}+1)}\left(\begin{array}{ccc}
j_{f} & m & j_{i}\\
\frac{1}{2} & 0 & -\frac{1}{2}\end{array}\right)\pi(l_{f},m,l_{i}),\label{ang-coeff}\end{equation}
 with

\begin{equation}
\pi(l_{f},m,l_{i})=\left\{ \begin{array}{c}
\begin{array}{cc}
1 & \mathrm{if}\: l_{f}+m+l_{i}\,\,\mathrm{even}\\
0 & \mathrm{otherwise}\end{array}\end{array}\right.\label{parity-rules}\end{equation}
 and $l$'s being the orbital angular momentum quantum numbers. When
$kr$ is sufficiently small, the spherical Bessel function $\mathtt{j}_{n}(kr)$
is approximated as

\begin{equation}
\mathtt{j}_{n}(kr)\approx\frac{(kr)^{n}}{(2n+1)!!}=\frac{(kr)^{n}}{1\cdot3\cdot5\cdot\cdot\cdot\cdot(2n+1)}.\label{bessel}\end{equation}

\section{\label{res}Results and discussions}

The magnetic (M1) and electric-dipole (E1) transition matrix elements
of Yb are computed using $37s33p28d12f5g$ GTOs with $\alpha_{0}=0.00525$
and $\beta=2.73$ (geometrical basis with $\alpha_{i}=\alpha_{0}\beta^{i-1}$).
{[}High lying unoccupied orbitals are not included (kept frozen) in
CC calculations.] The reference space for excitation energy and associated
properties is constructed by allocating $6s$ valence electrons of
Yb among $6s7s8s6p7p5d6d$ valence orbitals in all possible ways.
The basis and reference space used in this calculation is exactly
same as that employed in an earlier communication by one of the author
\cite{malaya-yb} for transition energies, ionization potential and
hyperfine matrix element calculations. We have considered that the
nucleus has a finite structure and is described by the two parameter
Fermi nuclear distribution \begin{equation}
\rho=\frac{\rho_{0}}{1+\exp((r-c)/a)}\,,\label{fermi-nucl}\end{equation}
 where the parameter $c$ is the half charge radius and $a$ is related
to skin thickness, defined as the interval of the nuclear thickness
in which the nuclear charge density falls from near one to near zero.
The energy levels of Yb and $\mathrm{Yb}^{+}$ are not reported here
as those have already appeared in the previous work \cite{malaya-yb}.
The magnetic-dipole transition matrix elements in Yb computed using
MR-FSCC method agree well with experiment and with other available
theoretical calculations (see Table \ref{tab1}.) The present result
for $\left|A(M1)\right|$ for $6s^{2}\,^{1}S_{0}\longrightarrow6s5d\,^{3}D_{1}$
transition differs by less than one percent ($<1\%$) from the experimental
value. Our calculation further shows that the major contribution to
$\left|A(M1)\right|$ comes from $S^{(0,1)}$ ($S^{(0,2)}$ contribution
is only 1\%). At this juncture, we emphasize that the random phase
approximation (RPA) and the second order multi-reference many-body
perturbation theory (MR-MBPT(2)) estimate this quantity ($\left|A(M1)\right|$)
to be $0.68\times10^{-4}\mu_{B}$ and $0.98\times10^{-4}\mu_{B}$
respectively. These large deviations ($\sim49\%$ for RPA and $\sim25\%$
for MBPT) in the perturbative estimate clearly demonstrates the importance
of higher order correlation effects.

\begin{table}

\caption{\label{tab1}Theoretical and experimental magnetic dipole transition
matrix elements (in Bohr magneton $\mu_{B}$) of Yb.}

~

\begin{centering}
\begin{tabular}{llll}
\hline 
Initial State &
Final State &
This work &
Expt./Theory \tabularnewline
\hline
\hline 
$6s^{2}\,^{1}S_{0}$ &
$6s5d\,^{3}D_{1}$&
$1.34\times10^{-4}$ &
$1.33\times10^{-4}$\cite{Budker} \tabularnewline
$6s6p\,^{3}P_{0}$ &
$6s6p\,^{1}P_{0}$ &
$0.12$ &
$0.13$\cite{Kimball} \tabularnewline
\hline
\end{tabular}
\par\end{centering}
\end{table}

In addition to the trasition matrix element $\left\langle 6s5d\,^{3}D_{1}\right|M1\left|6s^{2}\,^{1}S_{0}\right\rangle $,
we also report the $6s6p\,^{3}P_{0}\longrightarrow6s6p\,^{1}P_{1}$
$M1$ transition amplitude in Yb, which plays an important role in
the measurement of PNC induced electric-dipole amplitudes \cite{Kimball}.
We briefly outline its relevance as the details are available elsewhere
\cite{Kimball}. The PNC-induced electric-dipole transition amplitude
$A(E1)_{\mathrm{PNC}}$ is given by

\begin{equation}
{\displaystyle \begin{array}{ccc}
A(E1)_{\mathrm{PNC}} & = & \left\langle 6s6p\,^{1}P_{1}\right|ez\left|6s6p\,^{3}P_{0}\right\rangle \\
 & \approx & b{\displaystyle \frac{\left\langle 5d_{3/2}6s_{1/2}\right|H_{w}\left|5d_{3/2}6p_{1/2}\right\rangle }{\Delta E}}\\
 &  & \times\left\langle 6s5d\,^{3}D_{1}\right|ez\left|6s6p\,^{3}P_{0}\right\rangle \end{array}}\label{eq 17}\end{equation}
 where $H_{w}$ is the PNC weak interaction Hamiltonian in the non-relativistic
limit, $e$ is the electronic charge, $b$ is a coefficient that describes
the configuration mixing amplitude and angular mixing coefficient,
and $\Delta E$ is the energy separation between the $6s5d\,^{3}D_{1}$
and $6s6p\,^{1}P_{1}$ states \cite{Kimball}. The mixing coefficients
of the $6s5d\,^{3}D_{1}$ and $6s6p\,^{1}P_{1}$ states by the weak
interaction is given in Ref.\cite{Hw}. We have also determined the
matrix element $\left\langle 6s5d\,^{3}D_{1}\right|ez\left|6s6p\,^{3}P_{0}\right\rangle $
which turns out to be $2.52\,\mathrm{a.u.}$ This value provides a
step forward towards the determination of $A(E1)_{PNC}$ amplitude
in Yb and in the search of physics beyond the standard model.

\section{\label{conclusion}Conclusion}

We have computed the highly forbidden magnetic-dipole transition matrix
elements for $6s^{2}\,^{1}S_{0}\longrightarrow6s5d\,^{3}D_{1}$ and
$6s6p\,^{3}P_{0}\longrightarrow6s6p\,^{1}P_{1}$ transitions in Yb
using the Fock-space multi-reference coupled-cluster (MR-FSCC) method.
The values of the magnetic-dipole transition matrix elements presented
here are the most accurate theoretical estimates to date and are in
accord with the experimental value. We have also evaluated the $\left\langle 6s5d\,^{3}D_{1}\right|ez\left|6s6p\,^{3}P_{0}\right\rangle $
matrix element, which can be combined with other known quantities
to determine the PNC amplitude for the $6s^{2}\,^{1}S_{0}\longrightarrow6s5d\,^{3}D_{1}$
transition in atomic Yb. To our knowledge this the first time any
variant of coupled-cluster theory is applied to determine this quantity,
which is expected to be useful to experimentalists in this area and
in the search of any new physics beyond the standard model.

\begin{acknowledgments}
This work was partially supported by the National Science Foundation
and the Ohio State University (CS). RKC acknowledges the Department
of Science and Technology, India (grant SR/S1/PC-32/2005). C. S. acknowledges
Prof. B. P. Das for valuable discussions. We gracefully acknowledge
Prof. Russell Pitzer for his comments and criticism on the manuscript.
We sincerely acknowledge the constructive comments by the anonymous
referee. 
\end{acknowledgments}


\begin{thebibliography}{10}
\bibitem{Budker} J. E. Stalnaker, D. Budker, D. P. DeMille, S. J.
Freedman, V. V. Yashchuk, Phys. Rev. A \textbf{66}, 031403 (2002).

\bibitem{Hw} D. DeMille, Phys. Rev. Lett. \textbf{74}, 4165 (1995).

\bibitem{PNC-review}J. S. M. Ginges and V. V. Flaumbaum, \emph{Phys.
Rep.} \textbf{397}, 63 (2004).

\bibitem{Hv} K. F. Freed, in \emph{Lecture Notes in Chemistry}, edited
by U. Kaldor \textbf{52}, 1, Springer-Verlag, Berlin (1989).

\bibitem{MRMP} K. Hirao, \emph{Int. J. Quant. Chem.} \textbf{S26},
517 (1992); H. Nakano, \emph{J. Chem. Phys.} \textbf{99}, 7983 (1993).

\bibitem{PT2} K. Andersson, P. A. Malmqvista and B. O. Roos, \emph{J.
Chem. Phys.} \textbf{96}, 1218 (1992).

\bibitem{Hvr} R. K. Chaudhuri and K. F. Freed, \emph{J. Chem. Phys.}
\textbf{122}, 204111 (2005).

\bibitem{Nakano} M. Niyajima, Y. Watanbe and H. Nakano, \emph{J.
Chem. Phys.} \textbf{126}, 044101 (2006).

\bibitem{Dzuba} V. A. Dzuba, V. V. Flambaum, M. V. Marchenko, Phys.
Rev. \textbf{A 68} 022506 (2003).

\bibitem{mrcc} D. Mukherjee, R. K. Moitra, A. Mukhopadhyay, Mol.
Phys. \textbf{30}, 1861 (1975).

\bibitem{Lindgren} I. Lindgren, \emph{Int. J. Quant. Chem.} \textbf{S12},
33 (1978).

\bibitem{Mukherjee} D. Mukherjee, \emph{Proc. Ind. Acad. Sci.}, \textbf{96},
145 (1986); \emph{Chem. Phys. Lett.}, \textbf{125} 207 (1986); \emph{Int.
J. Quantum. Chem.}, \textbf{S20}, 409 (1986).

\bibitem{Kaldor} U. Kaldor, \emph{Recent Advances in Coupled-Cluster
Methods}, p 125, Ed. Rodney J. Bartlett, World Scientific, Singapore
(1997) and references therein.

\bibitem{Pie} X. Li, P. Piecuch and J. Paldus, \emph{Chem. Phys.
Lett.} \textbf{224}, 267 (1994).

\bibitem{NIST} NIST URL : http://www.nist.gov.

\bibitem{Kimball} D. F. Kimball, Phys. Rev. A \textbf{63}, 052113
(2001).

\bibitem{Haque} A. Haque, D. Mukherjee, J. Chem. Phys. \textbf{80},
5058 (1984).

\bibitem{LindMukh} I. Lindgren, D. Mukherjee, \emph{Phys. Rep.} \textbf{151},
93 (1987).

\bibitem{Sinha} D. Sinha, S. K. Mukhopadhyay, R. Chaudhuri, D. Mukherjee
D, Chem. Phys. Lett. \textbf{154}, 544 (1989).

\bibitem{Pal} S. Pal, M. Rittby, R. J. Bartlett, D. Sinha, D. Mukherjee,
Chem. Phys. Lett. \textbf{137}, 273 (1987).

\bibitem{SEC} D. Mukherjee , Pramana \textbf{12} 203 (1979).

\bibitem{malaya-yb} M. K. Nayak and R. K. Chaudhuri, \emph{Euro.
Phys. J. D} \textbf{37}, 171 (2006). 
\end{thebibliography}
\end{document}